\documentclass[]{aa}
\usepackage{graphicx}
\usepackage[utf8]{inputenc}
\usepackage{color}

\begin{document}
\newcommand{\sss}{$^{28}$Si:$^{29}$Si:$^{30}$Si}
\newcommand{\hho}{H$_2$O~}
\newcommand{\CC}{$^{12}$C/$^{13}$C~}
\newcommand{\CO}{$^{12}$C$^{16}$O~}
\newcommand{\CCO}{$^{13}$C$^{16}$O~}
\newcommand{\CCCO}{$^{14}$C$^{16}$O~}
\newcommand{\COOO}{$^{12}$C$^{18}$O~}
\newcommand{\COO}{$^{12}$C$^{17}$O~}
\newcommand{\SiO}{$^{28}$Si$^{16}$O~}
\newcommand{\SSiO}{$^{29}$Si$^{16}$O~}
\newcommand{\SSSiO}{$^{30}$Si$^{16}$O~}
\newcommand{\Tef}{$T_{\rm eff}$~}

\newcommand{\fel}{$f_{\rm el}$~}
\newcommand{\vsini}{$v\sin~i$~}
\newcommand{\lala}{$\lambda\lambda$~}
\newcommand{\Vt}{$V_t$~}
\newcommand{\DD}{$D_0$~}
\newcommand{\logg}{$\log$~g~}

\newcommand{\mum}{$\mu$m~}
\newcommand{\HHO}{H$_2$O~}

\newcommand{\red}[1]{{\color{red} #1}}
\newcommand{\green}[1]{{\color{green} #1}}


\title{Analysis of first overtone bands of isotopologues of CO and SiO in stellar spectra}

\author{Ya. V. Pavlenko \inst{1,3} \and
Sergei N. Yurchenko \inst{2}
  \and Jonathan Tennyson \inst{2}}

\institute{%
Main Astronomical Observatory of NAS Ukraine, 27 Zabolotnoho,  Kyiv, 01137, Ukraine
\and
Department of Physics and Astronomy, University College London, London WC1E 6BT, United Kingdom
\and
Centre for Astrophysics Research, University of Hertfordshire, College Lane, Hatfield, AL10 9AB, United Kingdom}

\date{\today}

\abstract{
The first overtone ($\Delta v = 2$) bands of the monosubstituted isotopologues of CO at 2.3 $\mu$m in the spectrum
of Arcturus (K2 III), and of the monosubstituted isotopologues of SiO
at 4 \mum in the spectrum of the red giant HD196610 (M6 III) are modelled.
}{
To investigate problems involving the computation of the first overtone bands of isotopologues of CO and SiO
in spectra of late-type
stars  and to determine isotopic abundances.
}{
We use fits of theoretical synthetic spectra to the observed stellar molecular bands of CO and SiO to determine
abundances of isotopes of C, O and Si.
}{
Fits of synthetic spectra of the \CO first overtone bands
at 2.3 \mum computed with three available line lists  (Goorvitch, HITEMP2010 and Li {\it et al.})
to the observed spectrum of Arcturus provide the same carbon
abundance [C]=$-0.6$ and isotopic ratio of carbon \CC=10 $\pm$ 2.
However, the quality of fits to the observed spectrum differ for three line lists used.
Furthermore, the  derived oxygen isotopic ratio $^{16}$O/$^{18}$O = 2000 $\pm$ 500 is
larger than that known in the solar system where  $^{16}$O/$^{18}$O = 500.
The silicon isotopic ratio in the atmosphere of the red giant HD196610 is revised.
Using the ExoMol SiO line list with appropriate
statistical weights for the SiO isotopologues the `non-solar'
ratio $^{28}$Si:$^{29}$Si:$^{30}$Si = 0.86$\pm$0.03:0.12$\pm$0.02:0.02$\pm$0.01 is obtained.
}{
We found that a) the computed isotopic carbon and silicon
 ratios determined by the fits to the observed spectrum
depend on the adopted abundance of C and Si, respectively;
b) Correct treatment of the nuclear spin degeneracies parameter is of crucial importance for the
use of nowadays HITRAN/ExoMol line lists in the astrophysical computations.
}

\maketitle

\titlerunning{ExoCross}
\authorrunning{Ya. Pavlenko et. al.}


\section{Introduction}

 Carbon has 15 known isotopes, from $^{8}$C to $^{22}$C, of which $^{12}$C and $^{13}$C are stable.
When a main-sequence star enters the red-giant branch, 
some convective mixing processes occur in its interiors that carry nuclei affected by CN cycling from internal layers
to the surface, see \cite{sned91, char07}. Theory predicts a decrease
in the \CC ratio with respect to the main sequence value ($\sim$ 89 in the case of the Sun),
down to values in
the range 15 -- 30, depending on the initial mass and metallicity of
the star, see \cite{char07}. Subsequently, the carbon abundance in the outer layers of red giants
drops, and nitrogen increases.

There are three known stable isotopes of oxygen: $^{16}$O, $^{17}$O, and $^{18}$O.
Radioactive isotopes ranging from $^{11}$O to $^{26}$O have also been characterized, all short-lived.
If the stellar mass of a red giant does not exceed 1.3 M$_{\odot}$ , the $^{16}$O abundance remains
unaltered, while that of $^{18}$O is mildly reduced, see \cite{palm11}.
In higher-mass stars, the dominant energy production process is the CNO cycle,
which is a catalytic cycle that uses nuclei of carbon, nitrogen and oxygen as
intermediaries and in the end produces a helium nucleus as with the
proton-proton chain \citep{bohm92}. However, in all cases the carbon and oxygen isotopic ratios
depend on the initial stellar masses and abundances \citep{char07}, so these parameters are of
crutial importance for understanding stellar evolution.

Silicon has 23 known isotopes, with mass numbers ranging from 22 to 44. $^{28}$Si (the most abundant
isotope), $^{29}$Si, and $^{30}$Si are stable.
All three stable isotopes of silicon are produced in stars through the oxygen-burning process, with the most abundant $^{28}$Si being made as part of the $\alpha$-process.  Oxygen-burning in massive stars  larger than 10.3 M$_{\odot}$ is preceded by the neon-burning process, although $^{16}$O is lighter than neon; oxygen in stars of lower masses cannot ignite.
After completing the oxygen burning processes the core of a star is primarily composed of silicon and sulfur, see \cite{clay83,
hirs14, caug88}, and references therein.  At advanced stages of  massive star evolution the core oxygen-burning process is succeeded by the fusion of $^{28}$Si with $\alpha$-particles prior to a violent type II supernova event, see \cite{naka99}.

It is worth noting that carbon, oxygen and silicon isotopes form in very specific and different nuclear reactions.
Knowledge of the isotopic composition of stars is therefore important for our understanding of how the Universe
works, see \cite{roma17}.

Unfortunately,
determination of isotopic abundances of C, O, Si from analysis their absorption lines in the optical spectral region
is complicated by the small wavelength shifts of the corresponding lines and the effects of blending which increase significantly
with lower effective stellar temperatures. Fortunately, molecular line positions shift significantly with isotopic substitution
and the infrared bands of different isotopologues
in   stellar spectra can provide input for the analysis of  isotopic abundances.

Vibrational bands of CO and SiO isotopologues are observed in spectra of late-type
stars with the first overtone ($\Delta v = 2$) bands having similar
shapes. However, due to the larger mass of the SiO molecule its fundamental
rovibrational spectrum  is redder at 4 \mum with respect to the
bands of CO located at 2.3 \mum.

Furthermore, because of its large dissociation energy and the relatively
high abundance of its constituent atoms, carbon monoxide in
its several isotopic forms is observed in a variety of astrophysical
sources, including erupting supernovae, stellar photospheres, comets, and the interstellar
medium, see e.g., \cite{roma17},  \cite{iwag15}, \cite{bane18}, and references therein.

Silicon monoxide, SiO, is an important constituent of these  astrophysical environments.  The
millemetre-range maser lines of  SiO  have been extensively used in the astrophysical applications, see \cite{tobi19}, \cite{vlem11}, and \cite{yoon14} for more details.
 Thermal lines of SiO in the
sub-mm and mm spectral  ranges have also been studied extensively as a probe
of circumstellar dust formation, conspicuous axial symmetry and bipolar dynamics, see
\cite{gonz03} for AGB stars and \cite{vice16} for a variety of evolved stars, as well as references therein.

The fundamental band of SiO was identified in the spectra of M supergiants a long time ago \citep{knac69, rins82, woll73}.
The strong 4 \mum band of SiO appears in the  spectra of
cooler stars, at least up to the spectral type M6. The first overtone band of SiO forms notable structures in the observed spectra of late-type stars, see \cite{evan19}, and \cite{pavl19} for more details. Furthermore, some recent theoretical studies of hot and dense atmospheres of exoplanets suggest  the possible presence of significant amounts of SiO and other silicates \citep{scha12, tine18}.

 In this work we obtain isotopic ratios for carbon, oxygen and silicon by analyzing spectra of two cool stars,
namely, Arcturus (K2 III) and HD196610 (M6 III).  Strictly speaking, we use Arcturus as a template star with known fundamental parameter to fit our synthetic spectra computed with the different line lists of various isotopologues of CO
to the observed stellar spectra.
In other words, a secondary aim of this paper is to compare  different CO and SiO line lists for the analysis of the isotopic abundance of stellar atmospheres and to demonstrate the importance of the consistent treatment of the molecular statistical weights when computing molecular opacity and partition functions.

\section{Line lists}

In the following we analyse different line lists for monosubstituted isotopologues of CO and SiO available in the literature: the three line lists for CO by  \cite{goor94}, HITEMP2010 \citep{roth10} and HITRAN2018 by \cite{li2015} and an SiO line list by \cite{bart13}.

\subsection{CO by \cite{goor94}}

\cite{goor94} computed carbon monoxide line lists for seven isotopologues, namely $^{12}$C$^{16}$O, $^{13}$C$^{16}$O,
$^{12}$C$^{17}$O, $^{12}$C$^{18}$O, $^{13}$C$^{18}$O, $^{14}$C$^{16}$O, and $^{13}$C$^{17}$O, containing
rotation-vibration transitions of the fundamental, first, and second overtone bands up to $v$ = 20 and $J$ = 149, as well
as pure rotational transitions for up to $v$ = 5 and $J$ = 60, of the ground  electronic state $X$~$^1 \Sigma^+$ .
The transitions dataset provided by \cite{goor94} comes with
the transition frequencies, the lower state energies, the Einstein A-values, the R-values, the expectation value of the
effective dipole moment operator, and the quantum numbers of each transition.

\subsection{CO in HITEMP2010}

The HITEMP2010 list list for CO isotopologues \citep{roth10} is in fact, the modified CO line list from HITEMP1995 in which
the data were assembled from the line list of \cite{goor94}.
 These changes are described by \cite{roth10}  as ``The only exception is
where there have been updates to the line parameters in
HITRAN for carbon monoxide after 1995; in these cases
the improved values were adopted for the HITEMP2010
compilation.'' Nevertheless, we were interested to see whether these
updates of the CO line list would change the results of the carbon isotope determination.

\subsection{CO by \cite{li2015}}

\cite{li2015} computed new extensive rovibrational line lists for nine isotopologues \CO,
\CCO, \COO, \CCCO, $^{13}$C$^{17}$O, $^{13}$C$^{18}$O, $^{14}$C$^{16}$O, $^{14}$C$^{17}$O, and $^{14}C^{18}$O in the
ground electronic state with $v <$ 42,
$ \Delta v <$ 12, and J $<$ 151. These line lists were subsequently included in
 HITRAN2016  \citep{hitr18}.
The line intensity and position calculations were computed using a newly determined piece-wise dipole moment function (DMF) in conjunction with wavefunctions calculated using an empirically determined potential energy function by \cite{coxo13}. A direct fit method that simultaneously fits all the reliable experimental rovibrational intensities was used to construct the DMF of CO at internuclear distances
near equilibrium.   \cite{li2015} claimed that molecular constants used to construct their line list, including those generated
by \cite{coxo04}, are accurate enough to reproduce all relevant spectroscopic line positions to the same degree of precision as that achieved in the direct potential fit.

\subsection{SiO by \cite{bart13}}
The EBJT line lists for silicon monoxide were computed by \cite{bart13} as part of the ExoMol project \citep{jt528} for the main isotopologue, \SiO, and for four monosubstituted isotopologues (\SSiO, \SSSiO, $^{28}$Si$^{18}$O and $^{28}$Si$^{17}$O), in their ground electronic states. These line lists
are suitable for high temperatures (up to 9000 K), including those relevant to exoplanetary
atmospheres, cool stars and sunspots. \cite{bart13} used a combination of empirical and ab initio methods: the
potential energy curves were determined to high accuracy by fitting to extensive data from
the analysis of both laboratory and sunspot spectra; a high-quality ab initio dipole moment
curves were calculated at the large basis set, with account of multireference configuration interactions.

\subsection{`Physics' and `astrophysics' conventions of the nuclear statistical weights and partition functions }

Two main conventions for definition of the nuclear statistical weights and associated partition functions $Q$ exist: the `physics' and `astrophysics' conventions. Their use requires  special care when using line lists for astrophysical applications. Let us consider an absorption coefficient  (line intensity) $I_{{\rm f}\gets {\rm i}}$ for a transition from an initial ${\rm i}$ to a final state f as given  in SI units, see \cite{15Bernath.book}:
\begin{equation}\label{e:I}
I(\mathrm{f}\leftarrow \mathrm{i})=\frac{e^2}{4 \varepsilon_0 m_{\rm e}} \frac{ g_{\rm i} f_{\rm if}  \, \mathrm{e}^{-E_{\mathrm{i}} / k T}}{Q}\left[1-\exp \left(-h c \tilde{\nu}_{\mathrm{if}} / k T\right)\right] ,
\end{equation}
where $f_{\rm if}$ is an oscillator strength, for the transition from the state i with energy $E_{\rm i}$, in thermal equilibrium at the temperature $T$, to the state f with energy $E_{\rm f}$, $\tilde{\nu}_{\rm if}$ as the transition wavenumber, $h c \tilde{\nu}_{\rm if}=E_{\mathrm{f}}-E_{\mathrm{i}}$, $m_{\rm e}$ is the electron mass, $e$ is the electron charge, $k$ is the Boltzmann constant, $g_{\rm i}$ is the degeneracy of the (lower) state i, $\varepsilon_{0}$ is the permittivity of free space and, finally, $Q$ is the partition function, given by
\begin{equation}\label{e:Q}
 Q=\sum_{i} g_i  \mathrm{e}^{-E_i/ k T}.
\end{equation}
The  statistical factor $g_i$ according to the `astrophysics' convention, commonly adopted by  astronomers, is given by
\begin{equation}\label{e:g_i:astron}
g_i = (2J_i+1)\, \frac{g_{i}^{(\rm ns)}}{\bar{g}^{(\rm ns)}},
\end{equation}
where  $g_i^{(\rm ns)}$ is the nuclear spin degeneracy of the state in question and $\bar{g}^{(\rm ns)}$ is the `total' nuclear spin factor. Examples of the partition functions $Q$ computed using this convention include works by \cite{81Irwin.partfunc} and \cite{84SaTaxx.partfunc}.
According to the `physics' convention, the statistical weight $g_i$ is the total degeneracy of the state $i$:
\begin{equation}\label{__g__}
g_i = (2J_i+1)\, g_{i}^{(\rm ns)}.
\end{equation}
This definition is adopted by major databases such as HITRAN \citep{jt692}, ExoMol \citep{jt631} and JANAF \citep{janaf}.
As one can see from Eqs.~(\ref{e:Q}) and (\ref{e:g_i:astron}), the factor $\bar{g}^{(\rm ns)}$ appears both in the numerator  and denominator of Eq.~\eqref{e:I} and thus cancels, i.e. both conventions give identical intensities, providing their consistent usage both in $Q$ and $I(\mathrm{f}\leftarrow \mathrm{i})$. However, this difference is a common source of errors when the conventions  used for the intensities (oscillator strength) and partition functions   are  mixed up, leading to non-physical results.


To convert between conventions:

\begin{equation}
\label{e:gf}
g_{\rm i} f_{\rm if}^{\rm (astro)} = \frac{{g_{\rm i} f_{\rm if}}^{\rm (phys)}}{\bar{g}^{(\rm ns)}}.
\end{equation}

As a very typical illustration, the nuclear spin degeneracies of the ortho and para states of water are $g_{\rm ortho}^{(\rm ns)} = $ 3 and $g_{\rm para}^{(\rm ns)} = $ 1. Thus, the corresponding `astrophysics' spin factors are 3/4 and 1/4 respectively (4 is the sum of $g_{\rm ortho}^{(\rm ns)}$ and $g_{\rm ortho}^{(\rm ns)}$), while the ``physics'' spin factors coincide with the nuclear spin degeneracies $g_{\rm ortho}^{(\rm ns)}$ and $g_{\rm para}^{(\rm ns)}$. Hence, the `astrophysics' and `physics' partition functions are related as
$$
\frac{Q^{\rm (phys)}}{Q^{\rm (astro)}} = 4.
$$
The wrong usage of the conventions can lead to the intensities being too strong or too weak by the factor of 4.

\begin{table}
\caption{Carbon, oxygen and silicon atom nuclear spin degeneracy parameters}
\label{_gs}
\centering
\begin{tabular}{ccc}
\hline\hline
Atom     &  $I$  & $2I+1$ \\ \hline
$^{12}$C & 0    &  1  \\
$^{13}$C & 1/2  &  2  \\
$^{14}$C & 0    &  1  \\
$^{16}$O & 0    &  1  \\
$^{17}$O & 5/2  &  6  \\
$^{18}$O & 0    &  1  \\
$^{28}$Si & 0   &  1  \\
$^{29}$Si & 1/2 &  2  \\
$^{30}$Si & 0   &  1  \\
\hline\hline

\end{tabular}
\end{table}

The nuclear spin degeneracy of a heteronuclear diatomic
molecule $AB$ molecule, $g \equiv g_{\rm if}$, is the same for all states and is given by
\begin{equation}
\label{e:gs}
g = (2\,I(A)+1)(2\,I(B)+1)
\end{equation}
where $I(X)$ is the nuclear spin of isotope $X$.

Turning to CO, \CO has zero nuclear spin degeneracy so in this case the factor is
unity meaning that `physics' and `astronomy' conventions are the same. The nuclear spin degeneracy factor of \CCO is 2
and therefore mixing different conventions in partition functions and oscillator strength would lead to a factor
of two difference.

For CO,
the \cite{goor94} line lists were computed in the framework of the `astrophysics' convention while those of
\cite{li2015} and HITEMP \citep{roth10}, as well as the SiO line lists of \cite{bart13},
used the `physics' convention.
To remain within one system of definitions we
 recomputed the $gf$ provided by \cite{li2015},  HITEMP2010 and \cite{bart13}
using the `astrophysics' convention.

\section{Spectral models}

\subsection{Observed spectra}

For the analysis, the infrared spectrum of Arcturus obtained by \cite{hink95} was used.
The resolution of the spectrum is of order 100~000 with the spectral lines broadened by the rotation
(\vsini = 1.5 -- 3 km/s) and macroturbulent motions in its atmosphere ($V_{macro} \sim$
4--6 km/s).

The HD196610 spectrum, taken from the IRTF stellar spectra library \citep{rayn09}, also
contains SiO bands. Despite much lower resolution ($R$ $\sim$ 2000), the structure of the first overtone bands is very evident.

\subsection{Opacities, model atmospheres}

For Arcturus we adopt \Tef = 4200 K, \logg =1.5, and a set of abundances determined by \cite{pete93}, with
$\log N$(Fe) = $-4.87$, $\log N$(C) = $-3.78$, $\log N$(O) = $-3.21$, $\log N$(Si) = $-4.59$.

 For HD196610, the computations were performed for \Tef = 3500 K. \cite{mcdo16} found from analysis
of CO and Na equivalent widths the metallicity [Fe/H] = $-0.3$ -- $-0.26$, however in their analysis they
used  lower \Tef by $\sim$ 250 K determined by \cite{cese13} and \cite{mcdo12}. Furthermore, they
determined for the star log g = 0.39,  see table 3 in \cite{mcdo16}.
To show possible effects of metallicity and gravity uncertainties on the
results of Si abundance and isotopic ratio determination we
carried out our analysis for the cases of [Fe/H] = 0 and $-0.3$, gravities logg = 1.0 and 0.5.
In this paper we adopt the ``solar'' scale of abundances
from \cite{ande89}: $\log N$(C) = $-3.48$, $\log N$(O)=$-3.11$, $\log N$(Si) = $-4.49$, but for iron we adopt
 $\log N$(Fe) = $-4.54$ from \cite{aspl09}.

Atomic lines from VALD3 \citep{ryab15}, as well as molecular lines
of H$_2$O, TiO, CrH, VO, CaH, \CO and \CCO were accounted in the
1D model atmosphere computations by \textsc{SAM12},
see more details in \cite{pavl03}.
Synthetic spectra
were computed for a 1D model atmosphere using the \textsc{Wita6} program \citep{pavl97} assuming Local Thermal Equilibrium (LTE).
Our model atmosphere and synthetic spectra were computed for the
microturbulent velocity of \Vt = 1.7 km/s.

\subsection{Synthetic spectra}

The synthetic spectra were computed using program
\textsc{WITA6} within a classical framework, e.g. LTE, hydro-static equilibrium and a
one-dimensional model atmosphere without sources and sinks of
energy.
Theoretical synthetic spectra were computed across the first overtone bands of CO (22500-22500 \AA)
and SiO (39000-42000\AA) with a wavelength step 0.05 \AA, opacities due to absorption by atoms, and \hho
were also accounted for.

\subsection{Fits to the observed spectra}

The best fit to the observed spectra was achieved  by the $\chi^2$ procedure described elsewhere \citep{pavl14}.
We give a few details here to aid understanding of the procedure. As part of the fit, the function
$$
S= \sum_{i=1}^{N} s_i^2
$$
is minimized, where $s_i= |F^{\rm obs}_i - F^{\rm comp}_i|$; $F^{\rm obs}_i$ and $F^{\rm comp}_i$ are the observed and computed
fluxes, respectively, and $N$ is the number of the wavelengths points  used in the minimisation procedure. In our analysis we omit some spectral ranges which contain artifacts provided by strong noise, telluric absorption, bad pixels, etc.
The minimisation factor $S$ is  computed on a 3D grid of radial velocity sets,
normalisation factors, and broadening parameters. The errors of the fit is evaluated as $\delta = \sum s_i/N$. In some of the plots,
the errors $\delta$ are indicated as error bars for $S$.

\section{Results}

\subsection{Identification of isotopic CO bands}

The CO first overtone bands system covers a wide
spectral range. Here we restrict our analysis at 22~360--25~000 \AA.
To verify the consistency of our results with \cite{pavl08co}, the carbon abundance and
carbon \CC isotopic ratio were determined for a narrower spectral range of 23~600--23~800~\AA,
which contains several molecular bands of the CO monosubstituted isotopologues.

\begin{figure}
   \centering
   \includegraphics[width=0.98\columnwidth]{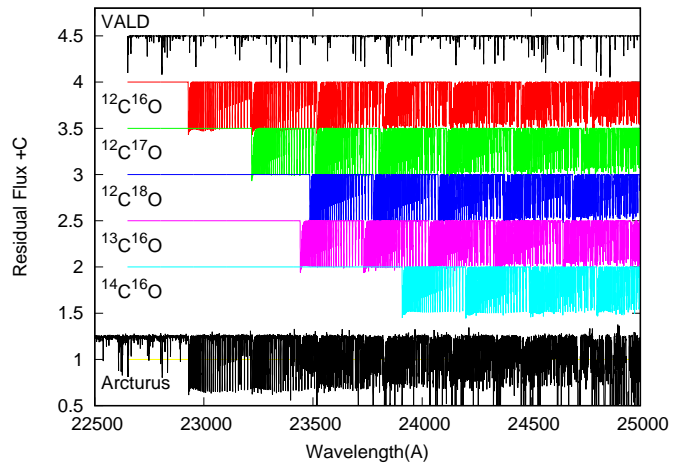}
\caption{\label{_ident} Positions of the CO isotopologue bands in
 the observed spectrum of Arcturus due to \cite{hink95}. Atomic absorption lines
 from VALD \citep{ryab15} across the spectral region are also shown.}
\end{figure}

The positions of the first overtone bands of the CO  isopologues in the observed spectrum of Arcturus is shown in Fig.~\ref{_ident}. We note
the distinct wavelengths differences in locations of \CO, \COOO, \CCO, \CCCO band heads,
with the \COO and \CO band heads overlapping. All molecular lists, except for \CCCO, were taken
from the HITEMP database, while the line list for \CCCO was taken from \cite{li2015}.

\subsection{Carbon abundance and \CC in Arcturus}
\label{_Cc}

It is worth noting that CO absorption dominates the spectrum of Arcturus across the fitted spectral region,
and that absorption by
\hho is very weak in this region. We therefore exclude  \hho from our analysis for the Arcturus data.

\begin{figure}
   \centering

   \includegraphics[width=0.98\columnwidth]{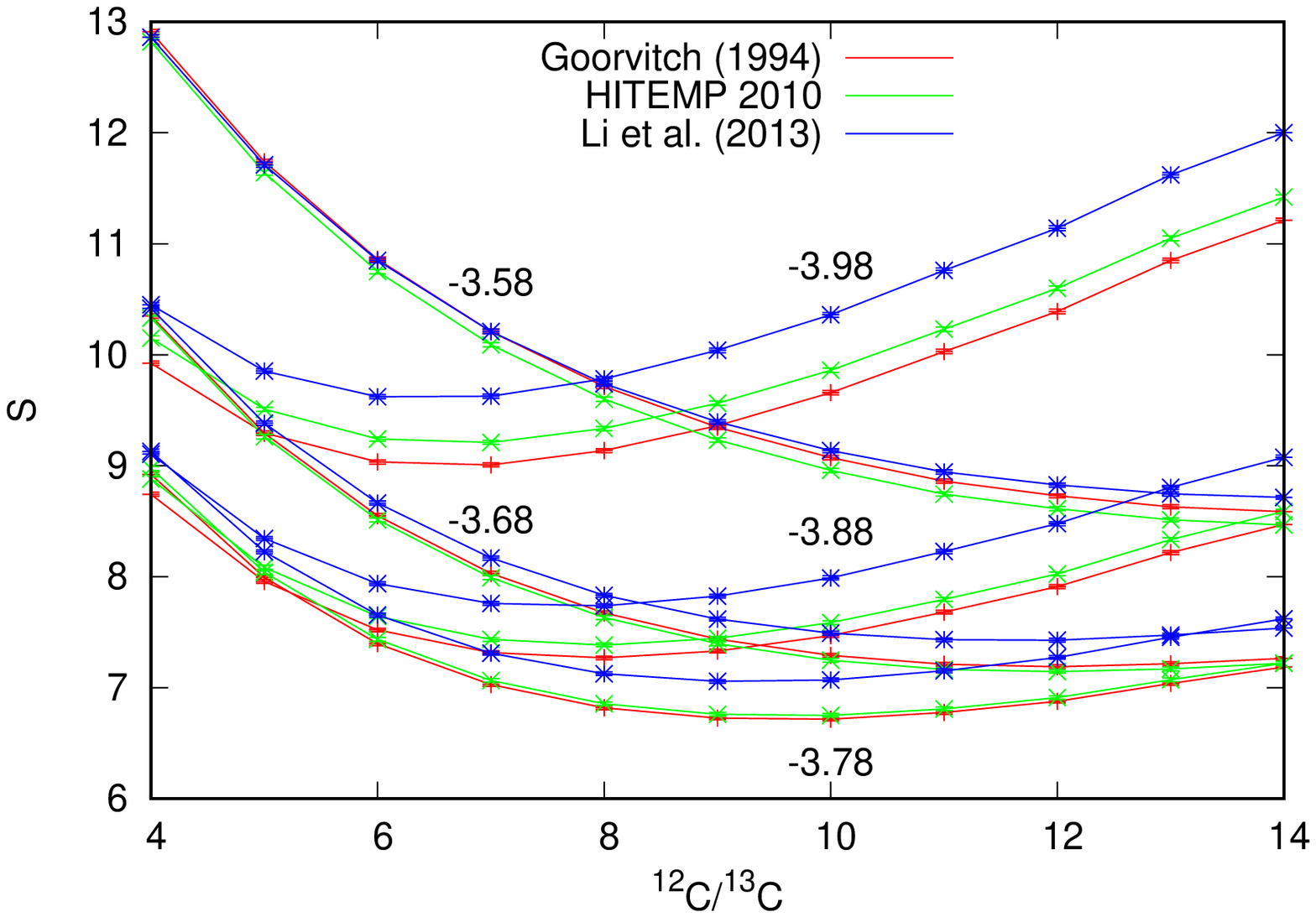}
   \includegraphics[width=0.98\columnwidth]{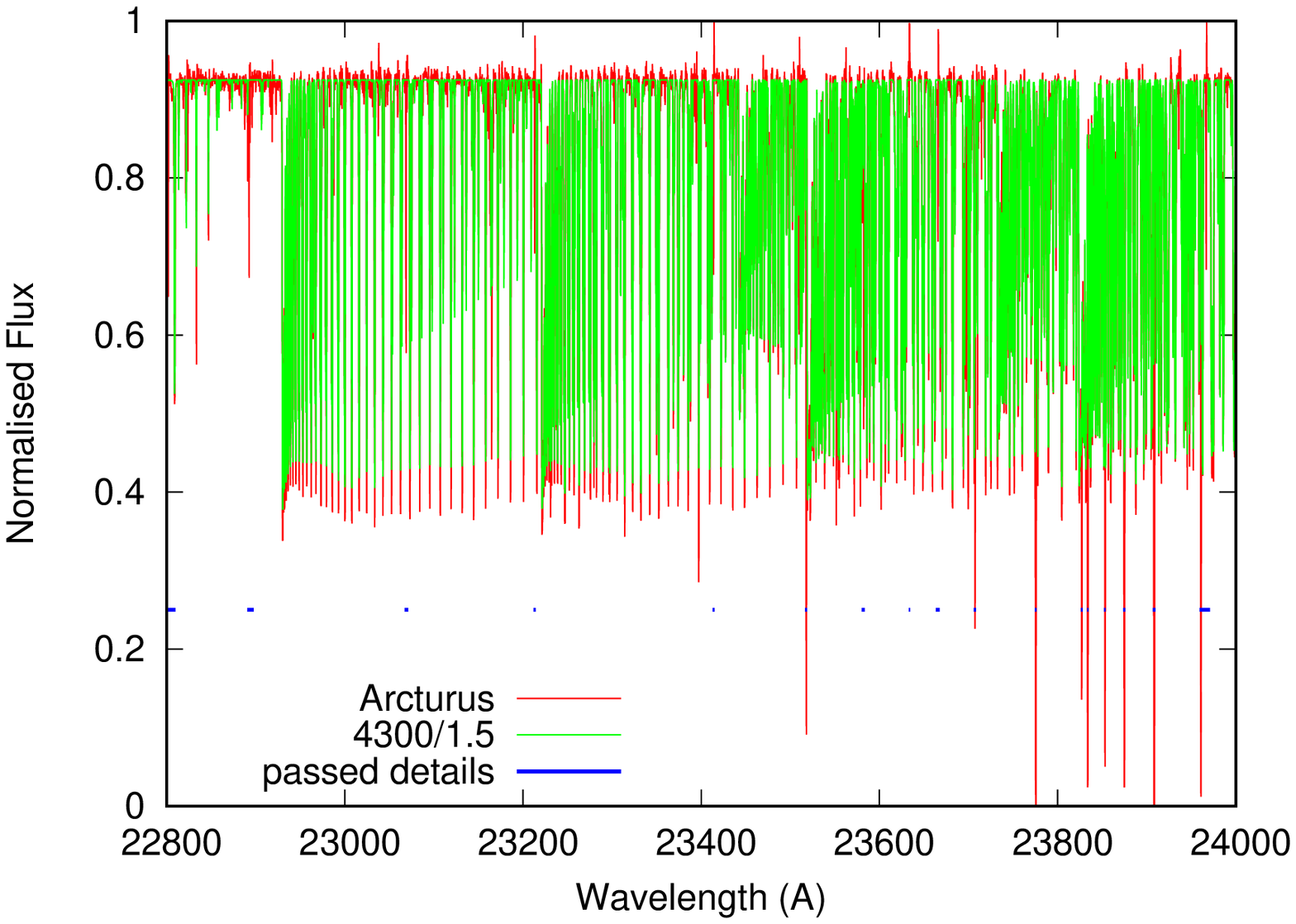}
\caption{\label{_fitsCCa} {\it Upper panel:}  Dependence of the minimisation parameter $S$ computed for
different line lists on the carbon abundance  in the atmosphere of
Arcturus. The adopted carbon abundance is shown above the correspondent curve. {\it Lower panel:} Best fit to the observed spectrum computed with $\log N$(C) = $-3.78$ and \CC =10.
 Spectral ranges containing artifacts which were missed by minimisation procedure are marked by thick blue lines.
}
\end{figure}

The carbon abundance and the carbon isotopic ratio were determined iteratively,
using line lists by \cite{goor94, roth10, li2015}. For any new carbon abundance
model the atmosphere of Arcturus was recomputed, resulting in $\log N$(C) = $-3.78
\pm$ 0.1 and \CC = 10 $\pm$ 2, respectively,  for all three line lists.  Our
fitting procedure ignores the `bad' spectral features, likely created by
telluric absorption lines, see \cite{hink95}. These features are marked by thick
blue lines in Fig. \ref{_fitsCCa} where the results of the fits to the observed
spectrum are shown. We note that our carbon abundance $\log N$(C) = $-3.78$
agrees within the error bar with the values found using a different procedure by
\cite{pete93}, \cite{ryde09}, and \cite{abia12}.

\subsection{$^{18}$O/$^{16}$O in Arcturus}

Molecular bands of \COOO could be observed in the modelled spectral region.
We estimated the ratio  $^{18}$O/$^{16}$O from a comparison of the relative strength of \CO and \COOO isotopologues.
To estimate the abundance of a given isotopologue we determine the relative number density of
the given species: $X_{^{12}{\rm C}^{18}{\rm O}} = N_{^{12}{\rm C}^{18}{\rm O}}/N_{\rm total}({\rm CO})$, where  $N_{\rm total}({\rm CO})$
is the total number density of all CO isotopologues. For the solar case, $X_{^{12}{\rm C}}$ = 0.989,
$X_{^{13}{\rm C}}$ = 0.011,
and $X_{^{16}{\rm O}}$ = 0.99762, $X_{^{18}{\rm O}}$ = 0.00200, see \cite{biev93}.

\begin{figure}
   \centering
   \includegraphics[width=0.98\columnwidth]{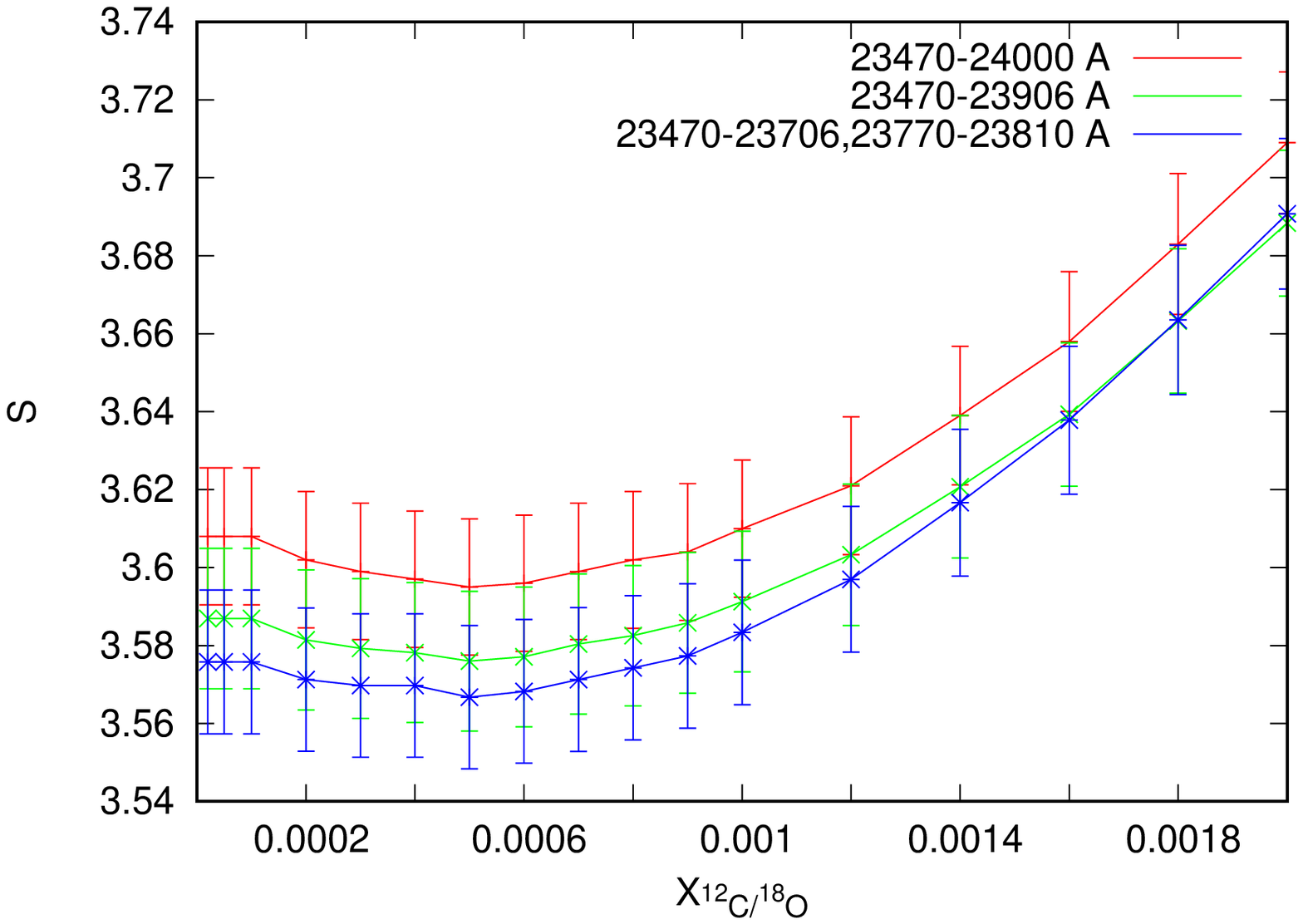}
   \includegraphics[width=0.98\columnwidth]{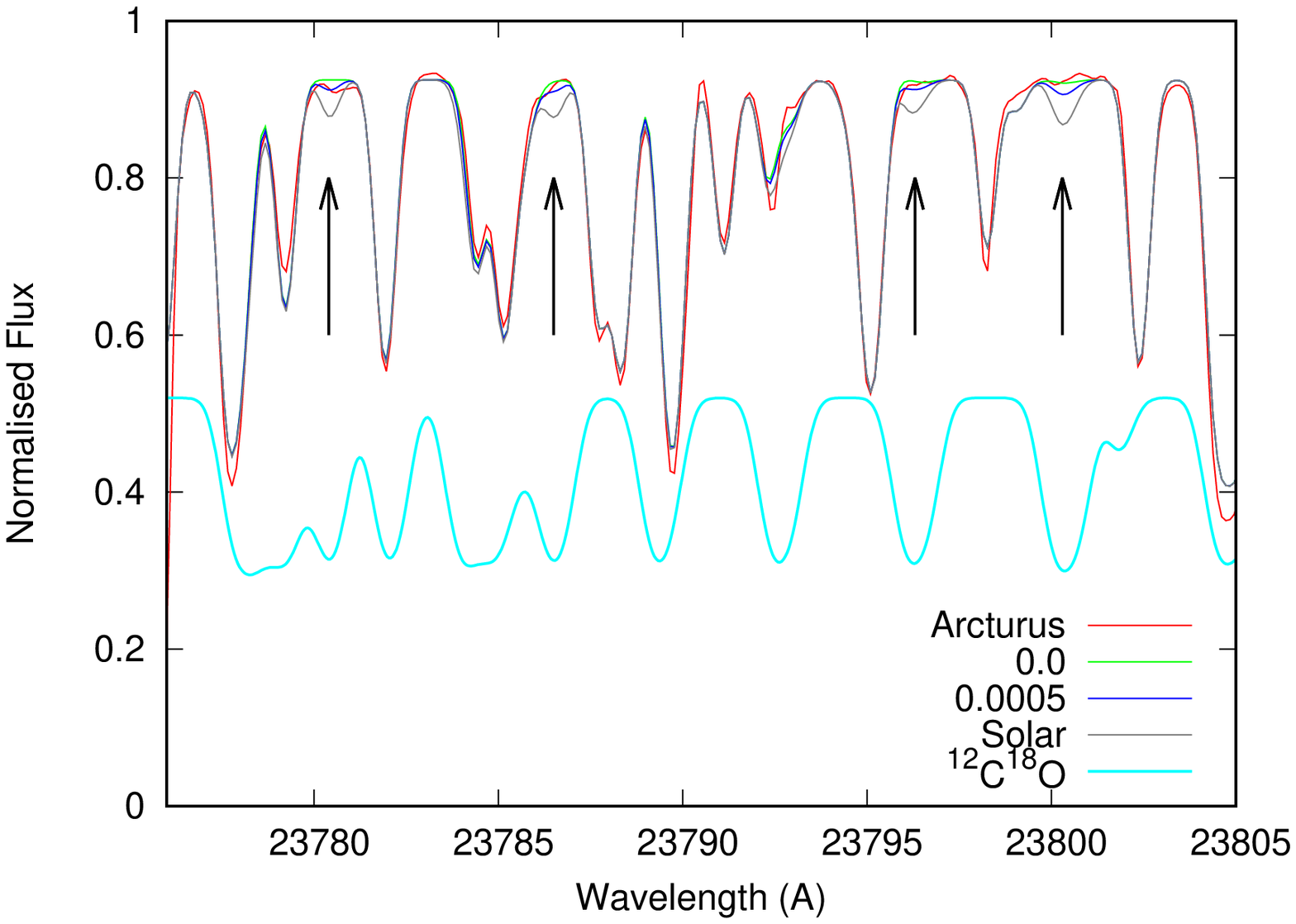}
\caption{\label{_fitsOO} {\it Upper panel:}  Dependence of the minimisation parameter computed for different
spectral ranges on $X_{^{12}{\rm C}^{18}{\rm O}}$.
{\it Lower panel:} Best fit to the observed spectrum computed for $X_{^{12}{\rm C}^{18}{\rm O}}$= 0.0001. Profile of the \COOO band is shown, in arbitrary units,  by a thick cyan line.  Vertical arrows mark
the \COOO features in the computed and observed spectrum.}
\end{figure}

Our  value of $X_{^{18}{\rm O}}$ =0.0005 $\pm$ 0.0004 was obtained at the min $S$ computed for different spectral ranges in the observed spectrum listed in
the upper panel of Fig. \ref{_fitsOO}. On the other hand, a synthetic spectrum generated with the `solar'
$X_{^{18}{\rm O}}$ gives  features which are notably absent from or too weak to be seen in the
observed spectrum, see the \COOO features marked by arrows at lower panel
of Fig.~\ref{_fitsOO}, which provides an  estimation `by eye'  of the lower abundance of $^{18}$O in the atmosphere of Arcturus compared to the Sun.

\subsection{$^{17}$O/$^{16}$O in Arcturus}

\begin{figure}
   \centering
   \includegraphics[width=0.98\columnwidth]{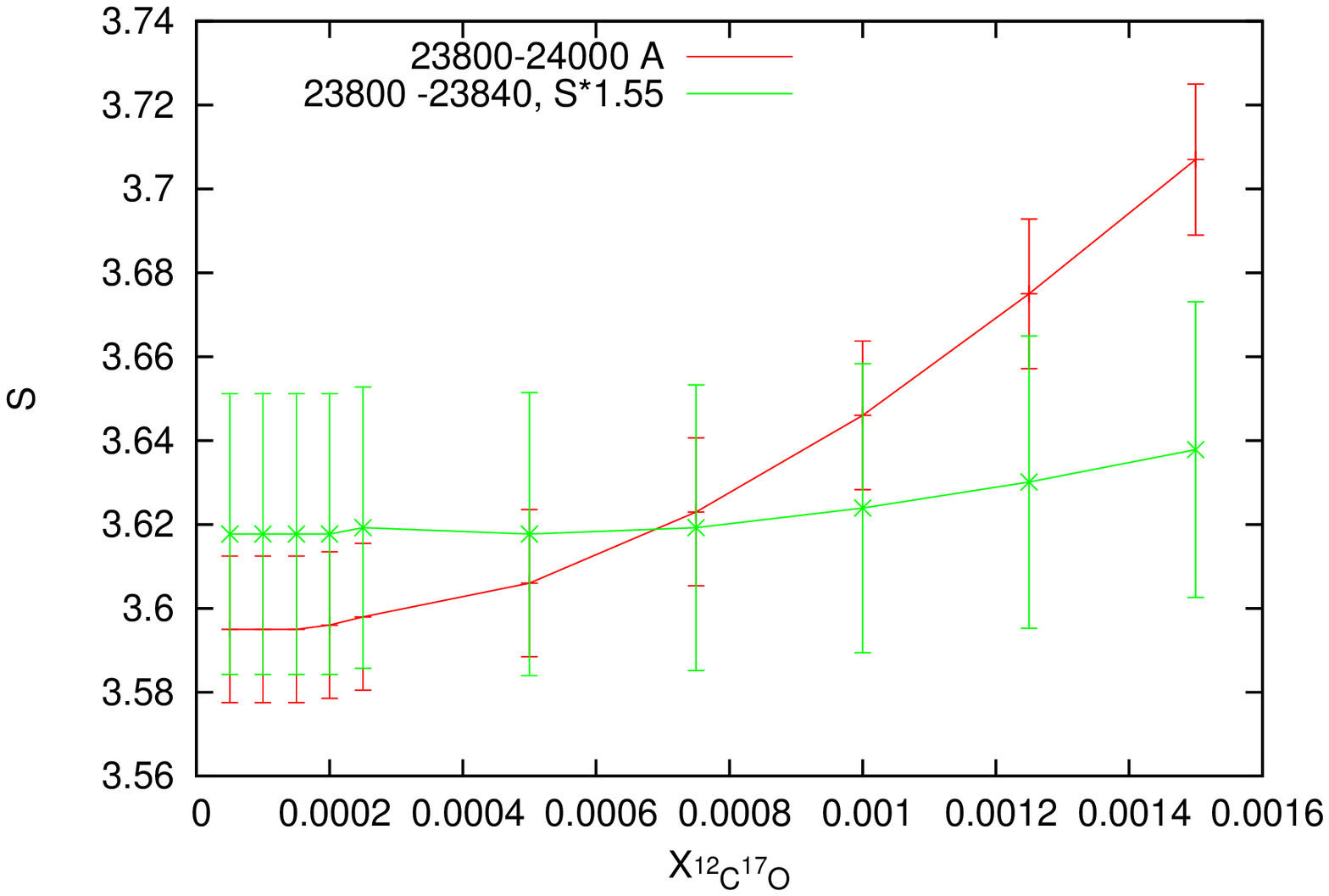}
   \includegraphics[width=0.98\columnwidth]{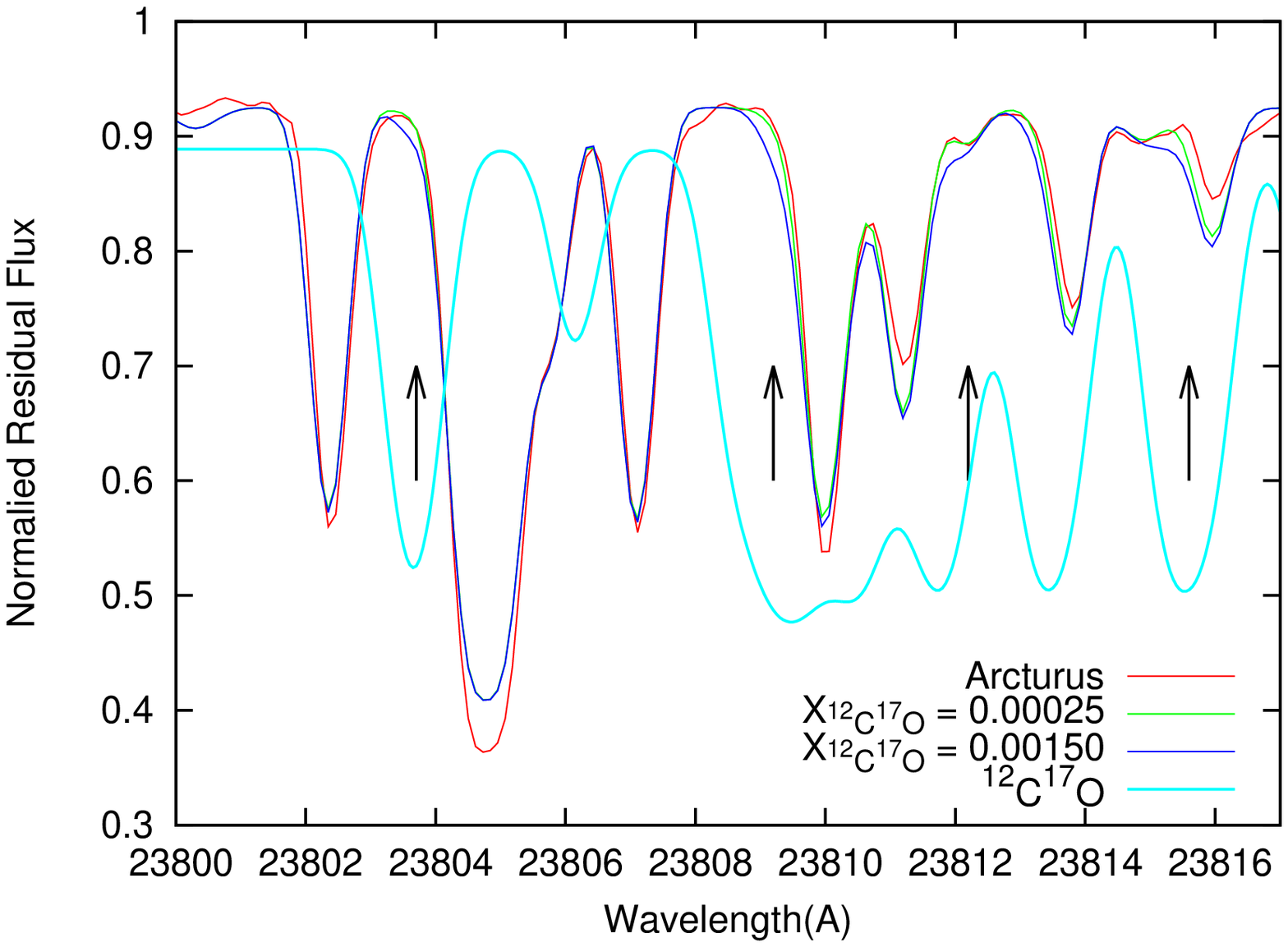}
\caption{\label{_fitsOOO} {\it Upper panel:}
 Dependence of the minimisation parameter computed for different
spectral ranges on $X_{^{12}{\rm C}^{17}{\rm O}}$.
{\it Lower panel:} Comparison of the best fit to the observed spectrum computed for $X_{^{12}{\rm C}^{17}{\rm O}}$= 0.00025 and 0.00150. Thick cyan line shows $^{12}$C$^{17}$O features at these wavelength, with the intensity arbitrary scale.
 Vertical arrows mark the positions of \COO features in the computed and observed spectrum.
}
\end{figure}

A similar procedure was used to estimate the ratio $^{17}$O/$^{16}$O in the
atmosphere of Arcturus, see Fig.~\ref{_fitsOOO};
for the Sun we have $X_{^{17}{\rm O}}$ = 0.00038 \citep{biev93}. We have some
evidence that the fraction  of $^{17}$O
is also lower in the Arcturus. However,
this estimation is harder than that for  $^{18}$O as determination of the
abundance of $^{17}$O in stellar atmospheres by  fitting to observed spectra is
hindered by  the coincidence between the wavelengths of the band heads of the
first overtone of {$^{12}$C$^{17}$O} band heads with those of the abundant
$^{12}$C$^{16}$O, see Fig. \ref{_ident}. Differences in these spectra increases
with the wavelength, however, at the longer wavelength, we observe more complex
spectral features created by a large number of CO bands. The pollution by
telluric lines also increases to the red spectral region. Taking into account
all these factors we may assume that $X_{^{12}{\rm C}^{17}{\rm O}} <$ 0.0002,
see upper panel of Fig. \ref{_fitsOOO}. However the formal error here is large
$\pm$ 0.0004.

\subsection{$^{14}$C/$^{12}$C in Arcturus}

\cite{goor94} and \cite{li2015} both computed line lists for the radioactive carbon monoxide isopologues $^{14}$C$^{16}$O,
$^{14}$C$^{17}$O and $^{14}$C$^{18}$O.
Radioactive carbon's isotope $^{14}$C can be formed in thermonuclear events on the stellar surface.  $^{14}$C has a half-life of 5,700 $\pm$ 40 years,\footnote{see http://www.nucleide.org/DDEP\_WG/Nuclides/C-14\_com.pdf} and
decays by emitting an electron and an electron antineutrino while one of the neutrons in the $^{14}$C
atom decays to a proton
into a stable (non-radioactive) isotope $^{14}$N atom. Unfortunately,
$^{14}$C$^{16}$O bands are shifted
redwards with respect to other CO isotopologues, which complicates their analysis due to
stronger blending by lines of other bands in this region, see Fig. \ref{_ident}.

\begin{figure}
   \centering
   \includegraphics[width=0.98\columnwidth]{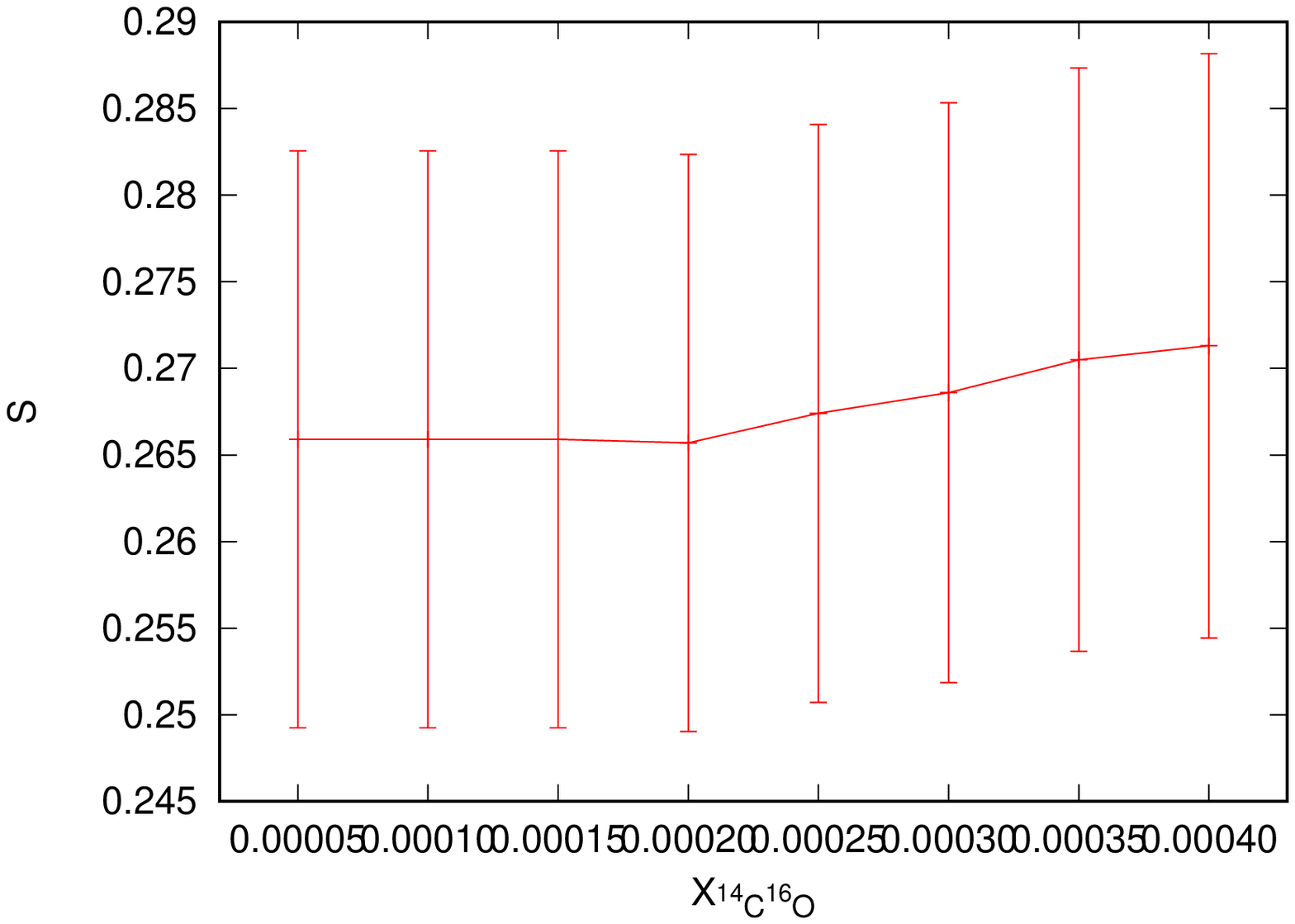}
   \includegraphics[width=0.98\columnwidth]{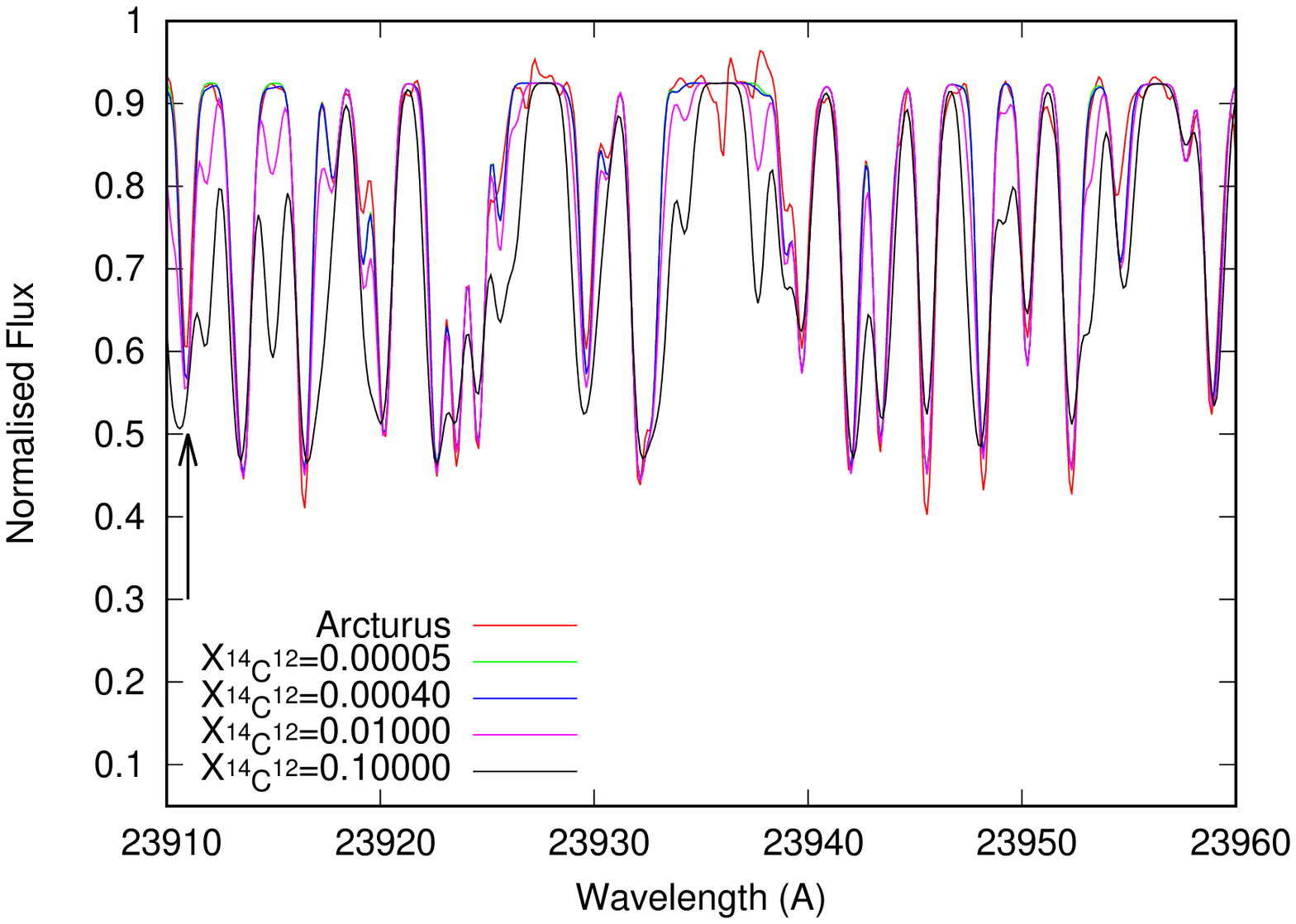}
\caption{\label{_CFO} {\it Upper panel:} Minimisation parameter $S$ computed for different $X_{^{14}{\rm C}^{16}{\rm O}}$.
{\it Lower panel:} Best fit to the observed spectrum computed for $X_{^{14}{\rm C}^{16}{\rm O}}$= 0.0005 and 0.00040.
Fits with values $X_{^{14}{\rm C}^{16}{\rm O}}$= 0.01 ad 0.01 are shown only for the evaluation purposes. The arrow at  23911 \AA\ marks
the position of the head of the $\Delta v = 2, v'=0$ band of the $^{14}{\rm C}^{16}$O molecule. Black thick lines mark
missed spectral features of telluric spectrum. }
\end{figure}

Naturally, most  $^{14}$C in the atmosphere of the star
should be in the form of the molecule $^{14}$C$^{16}$O.
We searched for $^{14}$C$^{16}$O in the spectrum of Arcturus across 23~900--24~000 \AA,
in the spectral range containing the head of $v'$ = 0 band of the molecule with negative result, no lines of  $^{14}$C$^{16}$O in the spectrum were found. From our modelling we may conclude that X$_{^{14}{\rm C}^{16}{\rm O}} < 0.00025$, see upper panel of Fig. \ref{_CFO} where the
position of the  $^{14}$C$^{16}$O band head computed for X$_{^{14}{\rm C}^{16}{\rm O}} =0.01$, is marked by the arrow in the lower panel of Fig. \ref{_CFO}
where we show the theoretical spectrum computed with X$_{^{14}{\rm C}^{16}{\rm O}} = 0.2$. The abundance of X$_{^{14}{\rm C}^{16}{\rm O}}$ seems to be  lower than our detectable limit or the molecule is completely absent  in the atmosphere of Arcturus.

\subsection{SiO first overtone bands in spectrum HD196610}

HD196610 is a variable red giant of \Tef =3500 $\pm$ 30 K,
see more details in \cite{pavl19}. In a recent paper \cite{pavl19} performed an initial fit to the observed
first overtone bands of SiO in the spectrum of the red giant HD196610 to determine isotopic ratios of silicon in its atmosphere. Unfortunately, \cite{pavl19} uses an inconsistent convention
for the nuclear spin statistical weights. We note that the factor $g_s = 2I+1$ =1 for the isotopologous \SiO and \SSSiO, while
for \SSiO $g_s$ = 2 (see Table \ref{_gs}). In this paper we revise the isotopic ratios of Si in the atmosphere of HD196610 by consistently using the `astrophysics' convention.

Furthermore, here we use a more sophisticated procedure than in \cite{pavl19} to determine the silicon isotopic ratios;
the same as was used for the analysis of the CO isotopolgues, see Section \ref{_Cc}. Here, the minimization of $S$ was
performed on a 2D grid of $X_{^{28}{\rm Si}^{16}{\rm O}}$
and $X_{^{29}{\rm Si}^{16}{\rm O}}$, at each point of the grid  $X_{^{30}{\rm Si}^{16}{\rm O}}$ =
$1-X_{^{29}{\rm Si}^{16}{\rm O}}-
X_{^{28}{\rm Si}^{16}{\rm O}}$. In this way we were able to determine the ratios for all three isotopologues of SiO.

It is worth noting that the determination of the min of $S$ was carried out iteratively, by varying the total abundance of silicon in the atmosphere of the star, which represents another improvement on the procedure of \cite{pavl19}.

\begin{figure}
   \centering
   \includegraphics[width=0.98\columnwidth]{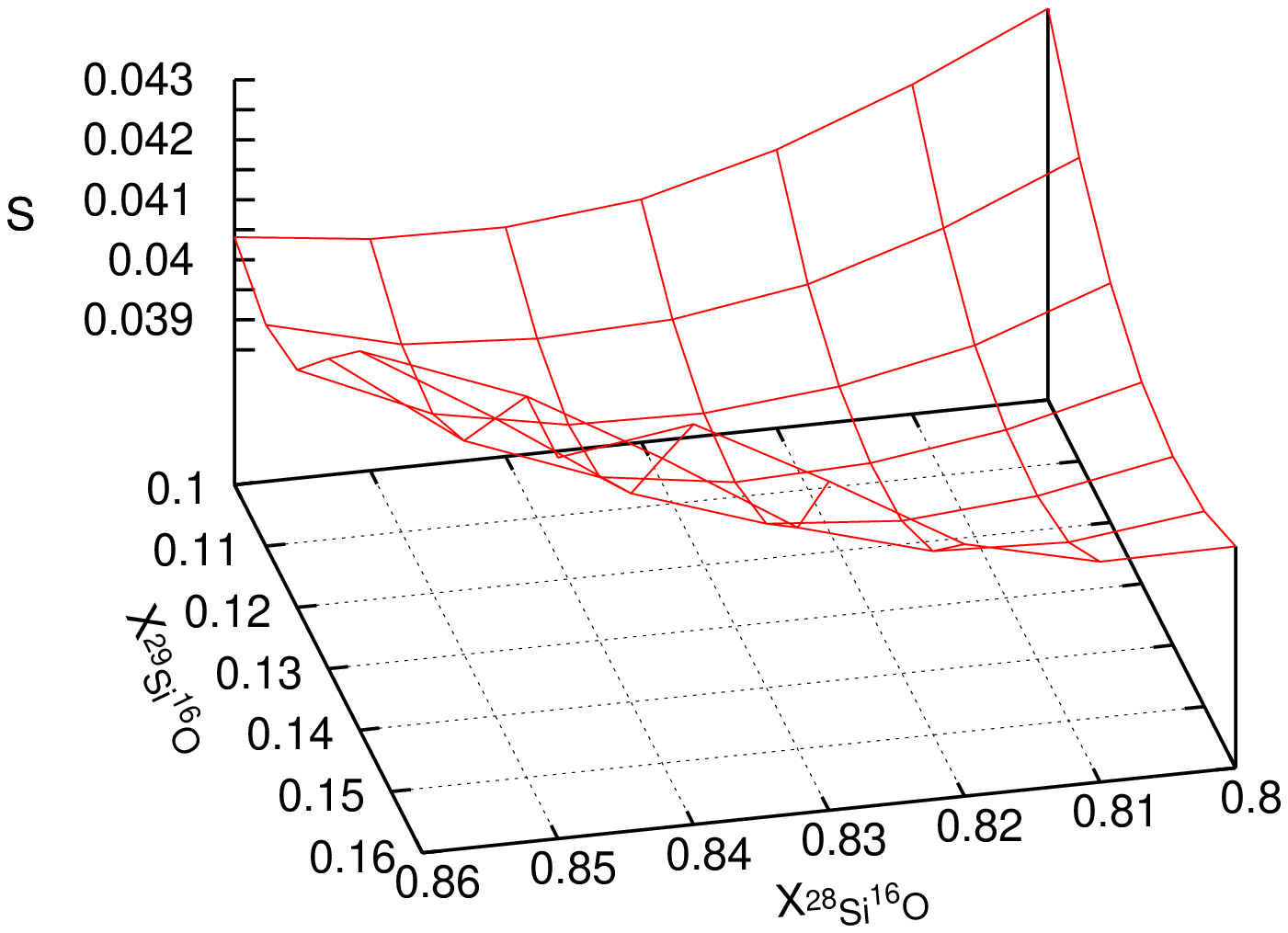}
   \includegraphics[width=0.98\columnwidth]{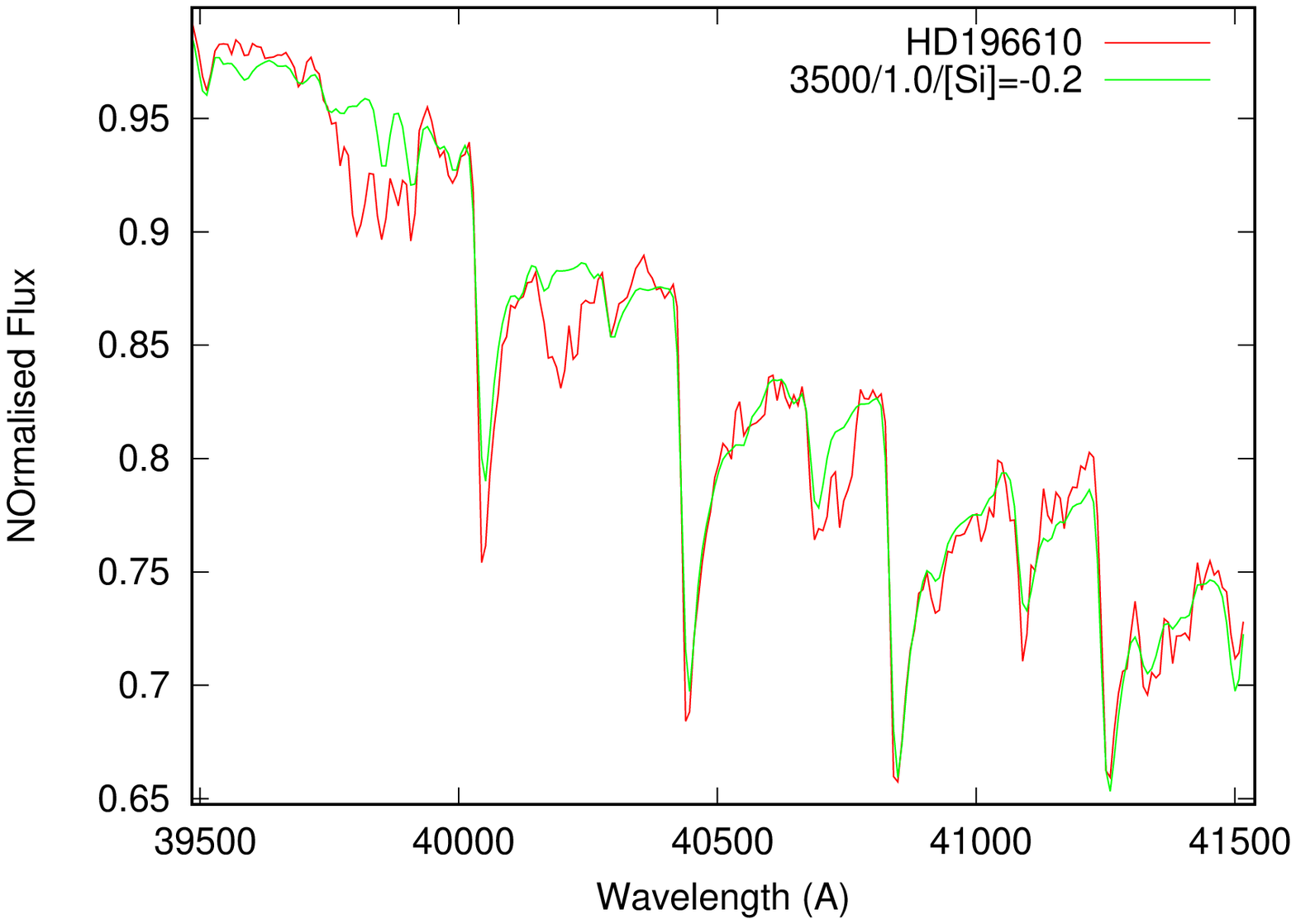}
\caption{\label{_sif}
{\it Upper panel:} Minimisation parameter $S$ computed on a grid of $X_{^{28}{\rm Si}^{16}{\rm O}}$ and $X_{^{29}{\rm Si}^{16}{\rm O}}$, here [Si] = $-0.2$.
{\it Lower panel:} Best fit to the observed spectrum  for [Si]=--0.2, $X_{^{28}{\rm Si}^{16}{\rm O}}$, $X_{^{29}{\rm Si}^{16}{\rm O}}$= 0.830 and 0.13, respectively. Other ''best solutions'' are given in the Table \ref{_sit}.
}
\end{figure}

A comparison of results is shown in Table \ref{_sit}. The upper panel of Fig. \ref{_sif} shows the best fit to the observed spectrum
containing the first overtone bands of SiO obtained with the parameters listed in Table \ref{_sit}.


\begin{table}
\caption{\label{_sit} Relative isotopologue abundances of Si obtained from the fits to the first
overtone bands of SiO observed in the atmosphere of HD196610. `Solar' isotopic ratios of Si \citep{biev93} are shown as well.}

\begin{tabular}{llllll}
\hline
\hline
\noalign{\smallskip}
 Star & [Si]   & $^{28}$Si     & $^{29}$Si     & $^{30}$Si & Refs.   \\
\noalign{\smallskip}
\hline
\noalign{\smallskip}
The Sun (5770/4.44/0.0) & 0.0 & 0.92         & 0.05          & 0.03 & [1]  \\
                        &              &               &        \\
HD196610 (3500/1.0/0.0) & 0.0    & 0.95         & 0.02          & 0.03  & [2] \\
                        &              &               &        \\
HD196610 (3500/1.0/0.0) &-0.2   &  0.83        & 0.13          & 0.04  & [3]  \\
                        &              &               &        \\
HD196610 (3500/1.0/-0.3)&-0.3   &  0.87        & 0.10          & 0.03  & [3]  \\
                        &              &               &        \\
HD196610 (3500/0.5/0.0) &-0.3   &  0.86        & 0.11          & 0.03  & [3]  \\
                        &              &               &        \\
HD196610 (3500/0.5/-0.3)&-0.4   &  0.89        & 0.10          & 0.01  & [3]  \\
                        &              &               &        \\

\hline\hline
\end{tabular}
\end{table}

Interestingly,  in the case of determination of silicon abundance together with
isotopic ratio  we obtain similar results for fits that
start with different Si abundances (and gravities).
Namely, we obtained [Si]=$-0.3 \pm$ 0.1, see Table \ref{_sit}.
We thus confirm the result of \cite{mcdo16}.
Accuracy of  $^{28}{\rm Si}^{16}{\rm O}$ abundance determination
 $X_{^{28}{\rm Si}^{16}{\rm O}}$ = 0.86 $\pm$ 0.03 is naturally limited by
 the low resolution of the observed spectra.
Nevertheless, we obtained here non-solar isotopic ratios for Si for all
input parameters tested.

As we see from a comparison with the results of \cite{pavl19}, the isotopic ratio computed with a consistent  convention for the statistical weights and partition functions changes the SiO isotopologue abundances significantly.

\section{Conclusion}

There are several line lists for the CO isotopologues available in the literature computed by different authors. In this paper we apply these line lists to analyse the observed spectrum of the  red giant Arcturus and show that all provide very similar results. However, the minimisation factor $S$ is lower for the line lists of \cite{goor94} and HITEMP2010 than for the newer \cite{li2015} line list which has been adopted by HITRAN, despite a more sophisticated approach used to generate the latter line list.
The detailed analysis of the differences is beyond the scope of this paper, here we note only, the difference in
the minimisation parameters $S$ obtained in the framework of the same procedure is clearly caused by the use of the different line lists.
On the other hand, the \CC\ ratios determined by the best fit to the observed spectrum are the same for all three line lists.

 In other words, a consistent usage of the statistical weight convention leads to the perfect agreement for the carbon
abundance and isotopic ratios between HITEMP2010 \citep{roth10} and the line lists of \cite{goor94}. However,
minimisation factor $S$
determined from the $\chi^2$ fit to observed spectrum for the cases of \cite{goor94} and \cite{roth10} lists is notably
smaller in
comparison with the case of \cite{li2015} line list.

The result of our carbon isotopic
determination from fits to the observed spectrum of Arcturus agrees,  at least within the error bars, with the previous
results by
\cite{bril94}, \cite{pavl03}, and \cite{abia12}. However,
we find  that the actual isotopologue abundance determined in the atmosphere of Arcturus depends on the
carbon abundance adopted, see Fig.~\ref{_fitsCCa}. We obtained the isotopic ratio \CC= 10 $\pm$ 2 and the
carbon abundance $\log N$(C) =$-3.78$
in the framework of our self-consistent approach, in which both \CC and $\log N$(C) vary in the process of finding the solution.

We find that the abundance of \COO  $X_{^{12}{\rm C}^{18}{\rm O}}$ = 0.0004 in the atmosphere of  Arcturus is lower than
that in the
Sun, i.e.,  $X_{^{12}{\rm C}^{18}{\rm O}}$ = 0.002, see \cite{biev93}. Arcturus is an older star than the Sun, so this may be interpreted as the evidence of a lower production of $^{18}$O at early epochs, see discussion in \cite{abia12} for more details.

 Using the fit of our synthetic spectra computed for the set of input parameters of HD196610,
e.g silicon abundances
and gravities we determined isotopic ratios  in the atmosphere of the star. The abundances and isotopic ratios were
determined in the framework  of self-consistent approach, when isotopic ratios and abundance of Si vary in
the process of determination of the best solution. In that way we found [Si] = $-0.3 \pm$ 0.1, the result agrees
with recent estimation of \cite{mcdo16}, obtained using very different procedure. Furthermore, we showed
that a) the Si abundance determination show rather marginal dependence on the input parameters, e.g.
abundances and gravity of the star, adopted in our work; b) we obtained ''non-solar'' isotopic ratio of Si
in atmosphere of HD196610, c) our estimations of $X_{^{28}{\rm Si}}$ = 0.86 $\pm$ 0.03 and $X_{^{29}{\rm Si}}$ = 0.86
$\pm$ 0.01, see Table  \ref{_sit}, shows rather marginal dependence on the adopted abundances and gravities,
as well.

However, we should note that these estimations were obtained by fits of our synthetic spectra to the spectra of the first overtone SiO bands obtained with rather low resolution. Naturally, more confidence in any analysis of the isotope ratios could be obtained using spectra of better quality;  and a more refined analysis should be followed by a detailed abundance analysis to reduce the number of free parameters.

Nevertheless, we demonstrate the importance of the consistent usage of spectroscopic data via a critical revision of the recent results by \cite{pavl19} on the  Si isotopic ratios. This leads to the difference of by a factor  more than 2 for \SSiO, from the nearly solar
ratio \sss = 0.92:0.05:0.03 to  the `non-solar' ratio 0.83:0.13:0.03. In this paper we also used a more advanced procedure to determine the silicon abundance and its isotopic ratios
in the framework of our self-consistent approach, in which both parameters vary in the iteratively
process of finding solution.

 And finally, it can be assumed with a high degree of confidence that, in the case of these two stars, we observe differences in the isotopic composition of oxygen and silicon between the times and places of their formation and the present day values. Indeed, the masses of both stars are 0.6  and 1.1  solar masses for HD196610 \citep{mcdo16} and Arcturus \citep{ayre77}, respectively. Definitely these masses are not high enough to initiate nucleosynthesis of silicon and oxygen. Detailed analysis of spacial distributions of isotopes in our Galaxy is of crutial importance and vital for modern astrophysics, but the problem is beyond the scope of this paper.

\section*{Acknowledgements}

This study was funded as part of the routine financing program for institutes of the National Academy of Sciences
of Ukraine.
Spectroscopic data calculated by the
ExoMol group (funded by ERC as part of the Advanced Investigator 267219 project and UK Science and Technology Research Council (STFC) No. ST/R000476/1), the SIMBAD database (CDS,
Strasbourg, France), and the Gaia spacecraft data (European Space Agency) were used. This study is based in part
on archival data obtained using the infrared telescope operated by the University of Hawaii under a cooperative
agreement with NASA. Authors would like to thank the SAO/NASA ADS team for the development and support
of this remarkable data system.  We thank the anonymous referee for a thorough review and we highly
appreciate the comments and suggestions, which significantly contributed
 to improving the quality of the paper.



\begin{thebibliography}{55}
\expandafter\ifx\csname natexlab\endcsname\relax\def\natexlab#1{#1}\fi

\bibitem[{{Abia} {et~al.}(2012){Abia}, {Palmerini}, {Busso}, \&
  {Cristallo}}]{abia12}
{Abia}, C., {Palmerini}, S., {Busso}, M., \& {Cristallo}, S. 2012, \aap, 548,
  A55

\bibitem[{{Anders} \& {Grevesse}(1989)}]{ande89}
{Anders}, E. \& {Grevesse}, N. 1989, \gca, 53, 197

\bibitem[{{Asplund} {et~al.}(2009){Asplund}, {Grevesse}, {Sauval}, \&
  {Scott}}]{aspl09}
{Asplund}, M., {Grevesse}, N., {Sauval}, A.~J., \& {Scott}, P. 2009, \araa, 47,
  481

\bibitem[{{Ayres} \& {Johnson}(1977)}]{ayre77}
{Ayres}, T.~R. \& {Johnson}, H.~R. 1977, \apj, 214, 410

\bibitem[{Banerjee {et~al.}(2018)Banerjee, Joshi, Evans, Srivastava, Ashok,
  Gehrz, Connelley, Geballe, Spyromilio, Rho, \& Roy}]{bane18}
Banerjee, D. P.~K., Joshi, V., Evans, A., {et~al.} 2018, Monthly Notices of the
  Royal Astronomical Society, 481, 806

\bibitem[{{Barton} {et~al.}(2013){Barton}, {Yurchenko}, \& {Tennyson}}]{bart13}
{Barton}, E.~J., {Yurchenko}, S.~N., \& {Tennyson}, J. 2013, \mnras, 434, 1469

\bibitem[{Bernath(2015)}]{15Bernath.book}
Bernath, P.~F. 2015, Spectra of Atoms and Molecules, 3rd edn. (Oxford
  University Press)

\bibitem[{{B{\"o}hm-Vitense}(1992)}]{bohm92}
{B{\"o}hm-Vitense}, E. 1992, {Introduction to Stellar Astrophysics}

\bibitem[{{Briley} {et~al.}(1994){Briley}, {Smith}, \& {Lambert}}]{bril94}
{Briley}, M.~M., {Smith}, V.~V., \& {Lambert}, D.~L. 1994, \apjl, 424, L119

\bibitem[{Caughlan \& Fowler(1988)}]{caug88}
Caughlan, G.~R. \& Fowler, W.~A. 1988, Atomic Data and Nuclear Data Tables, 40,
  283

\bibitem[{{Cesetti} {et~al.}(2013){Cesetti}, {Pizzella}, {Ivanov}, {Morelli},
  {Corsini}, \& {Dalla Bont{\`a}}}]{cese13}
{Cesetti}, M., {Pizzella}, A., {Ivanov}, V.~D., {et~al.} 2013, \aap, 549, A129

\bibitem[{{Charbonnel} \& {Zahn}(2007)}]{char07}
{Charbonnel}, C. \& {Zahn}, J.~P. 2007, \aap, 467, L15

\bibitem[{Chase \& et~al.(1985)}]{janaf}
Chase, M.~W. \& et~al. 1985, J. Phys. Chem. Ref. Data, 14

\bibitem[{{Clayton}(1983)}]{clay83}
{Clayton}, D.~D. 1983, {Principles of stellar evolution and nucleosynthesis}

\bibitem[{Coxon \& Hajigeorgiou(2004)}]{coxo04}
Coxon, J.~A. \& Hajigeorgiou, P. 2004, The Journal of chemical physics, 121,
  2992

\bibitem[{Coxon \& Hajigeorgiou(2013)}]{coxo13}
Coxon, J.~A. \& Hajigeorgiou, P. 2013, Journal of Quantitative Spectroscopy and
  Radiative Transfer, 116, 75–78

\bibitem[{{De Bievre} \& {Taylor}(1993)}]{biev93}
{De Bievre}, P. \& {Taylor}, P. 1993, International Journal of Mass
  Spectrometry and Zon Processes, 123, 149

\bibitem[{{de Vicente} {et~al.}(2016){de Vicente}, {Bujarrabal},
  {D{\'\i}az-Pulido}, @ARTICLE{2016A&A...589A..74D, author = {{de Vicente}, P.
  and {Bujarrabal}, V. and {D{\'\i}az-Pulido}, A. and {Albo}, C. and {Alcolea},
  J. and {Barcia}, A. and {Barbas}, L. and {Bola{\~n}o}, R. and {Colomer}, F.
  and {Diez}, M.~C. and {Gallego}, J.~D. and {G{\'o}mez-Gonz{\'a}lez}, J. and
  {L{\'o}pez-Fern{\'a}ndez}, I. and {L{\'o}pez-Fern{\'a}ndez}, J.~A. and
  {L{\'o}pez-P{\'e}rez}, J.~A. and {Malo}, I. and {Moreno}, A. and {Patino}, M.
  and {Serna}, J.~M. and {Tercero}, F. and {Vaquero}, B.}, title = "{$^{28}$SiO
  v = 0 J = 1-0 emission from evolved stars}", journal = {\aap}, keywords =
  {stars: AGB and post-AGB, circumstellar matter, radio lines: stars,
  Astrophysics - Astrophysics of Galaxies, Astrophysics - Solar and Stellar
  Astrophysics}, year = "2016", month = "May", volume = {589}, eid = {A74},
  pages = {A74}, doi = {10.1051/0004-6361/201527174}, archivePrefix = {arXiv},
  eprint = {1603.01163}, primaryClass = {astro-ph.GA}, adsurl =
  {https://ui.adsabs.harvard.edu/abs/2016A&A...589A..74D}, adsnote = {Provided
  by the SAO/NASA Astrophysics Data System} }~{Albo}, {Alcolea}, {Barcia},
  {Barbas}, {Bola{\~n}o}, {Colomer}, {Diez}, {Gallego},
  {G{\'o}mez-Gonz{\'a}lez}, {L{\'o}pez-Fern{\'a}ndez},
  {L{\'o}pez-Fern{\'a}ndez}, {L{\'o}pez-P{\'e}rez}, {Malo}, {Moreno}, {Patino},
  {Serna}, {Tercero}, \& {Vaquero}}]{vice16}
{de Vicente}, P., {Bujarrabal}, V., {D{\'\i}az-Pulido}, A., {et~al.} 2016,
  \aap, 589, A74

\bibitem[{{Evans} {et~al.}(2019){Evans}, {Pavlenko}, {Banerjee}, {Munari},
  {Gehrz}, {Woodward}, {Starrfield}, {Helton}, {Shahbandeh}, {Davis},
  {Dallaporta}, \& {Cherini}}]{evan19}
{Evans}, A., {Pavlenko}, Y.~V., {Banerjee}, D.~P.~K., {et~al.} 2019, \mnras,
  486, 3498

\bibitem[{Gamache {et~al.}(2017)Gamache, Roller, Lopes, Gordon, Rothman,
  Polyansky, Zobov, Kyuberis, Tennyson, Yurchenko, Cs\'asz\'ar, Furtenbacher,
  Huang, Schwenke, Lee, Drouin, Tashkun, Perevalov, \& Kochanov}]{jt692}
Gamache, R.~R., Roller, C., Lopes, E., {et~al.} 2017, J. Quant. Spectrosc.
  Radiat. Transf., 203, 70

\bibitem[{{Gonz{\'a}lez Delgado} {et~al.}(2003){Gonz{\'a}lez Delgado},
  {Olofsson}, {Kerschbaum}, {Sch{\"o}ier}, {Lindqvist}, \&
  {Groenewegen}}]{gonz03}
{Gonz{\'a}lez Delgado}, D., {Olofsson}, H., {Kerschbaum}, F., {et~al.} 2003,
  \aap, 411, 123

\bibitem[{{Goorvitch}(1994)}]{goor94}
{Goorvitch}, D. 1994, \apjs, 95, 535

\bibitem[{{Gordon} {et~al.}(2017){Gordon}, {Rothman}, {Hill}, {Kochanov},
  {Tan}, {Bernath}, {Birk}, {Boudon}, {Campargue}, {Chance}, {Drouin}, {Flaud},
  {Gamache}, {Hodges}, {Jacquemart}, {Perevalov}, {Perrin}, {Shine}, {Smith},
  {Tennyson}, {Toon}, {Tran}, {Tyuterev}, {Barbe}, {Cs{\'a}sz{\'a}r}, {Devi},
  {Furtenbacher}, {Harrison}, {Hartmann}, {Jolly}, {Johnson}, {Karman},
  {Kleiner}, {Kyuberis}, {Loos}, {Lyulin}, {Massie}, {Mikhailenko},
  {Moazzen-Ahmadi}, {M{\"u}ller}, {Naumenko}, {Nikitin}, {Polyansky}, {Rey},
  {Rotger}, {Sharpe}, {Sung}, {Starikova}, {Tashkun}, {Auwera}, {Wagner},
  {Wilzewski}, {Wcis{\l}o}, {Yu}, \& {Zak}}]{hitr18}
{Gordon}, I.~E., {Rothman}, L.~S., {Hill}, C., {et~al.} 2017, \jqsrt, 203, 3

\bibitem[{{Hinkle} {et~al.}(1995){Hinkle}, {Wallace}, \& {Livingston}}]{hink95}
{Hinkle}, K., {Wallace}, L., \& {Livingston}, W. 1995, \pasp, 107, 1042

\bibitem[{Hirschi(2014)}]{hirs14}
Hirschi, R. 2014, Astrophysics and Space Science Library, 157–198

\bibitem[{Irwin(1981)}]{81Irwin.partfunc}
Irwin, A.~W. 1981, ApJS, 45, 621

\bibitem[{{Iwagami} {et~al.}(2015){Iwagami}, {Hashimoto}, {Ohtsuki}, {Takagi},
  \& {Robert}}]{iwag15}
{Iwagami}, N., {Hashimoto}, G.~L., {Ohtsuki}, S., {Takagi}, S., \& {Robert}, S.
  2015, \planss, 113, 292

\bibitem[{{Knacke} {et~al.}(1969){Knacke}, {Gaustad}, {Gillett}, \&
  {Stein}}]{knac69}
{Knacke}, R.~F., {Gaustad}, J.~E., {Gillett}, F.~C., \& {Stein}, W.~A. 1969,
  \apjl, 155, L189

\bibitem[{{Li} {et~al.}(2015){Li}, {Gordon}, {Rothman}, {Tan}, {Hu}, {Kassi},
  {Campargue}, \& {Medvedev}}]{li2015}
{Li}, G., {Gordon}, I.~E., {Rothman}, L.~S., {et~al.} 2015, \apjs, 216, 15

\bibitem[{{McDonald} {et~al.}(2012){McDonald}, {Zijlstra}, \& {Boyer}}]{mcdo12}
{McDonald}, I., {Zijlstra}, A.~A., \& {Boyer}, M.~L. 2012, \mnras, 427, 343

\bibitem[{{McDonald} {et~al.}(2016){McDonald}, {Zijlstra}, {Sloan}, {Lagadec},
  {Johnson}, {Uttenthaler}, {Jones}, \& {Smith}}]{mcdo16}
{McDonald}, I., {Zijlstra}, A.~A., {Sloan}, G.~C., {et~al.} 2016, \mnras, 456,
  4542

\bibitem[{{Nakamura} {et~al.}(1999){Nakamura}, {Umeda}, {Nomoto}, {Thielemann},
  \& {Burrows}}]{naka99}
{Nakamura}, T., {Umeda}, H., {Nomoto}, K., {Thielemann}, F.-K., \& {Burrows},
  A. 1999, \apj, 517, 193

\bibitem[{{Palmerini} {et~al.}(2011){Palmerini}, {La Cognata}, {Cristallo}, \&
  {Busso}}]{palm11}
{Palmerini}, S., {La Cognata}, M., {Cristallo}, S., \& {Busso}, M. 2011, \apj,
  729, 3

\bibitem[{{Pavlenko}(1997)}]{pavl97}
{Pavlenko}, Y.~V. 1997, Astronomy Reports, 41, 537

\bibitem[{{Pavlenko}(2003)}]{pavl03}
{Pavlenko}, Y.~V. 2003, Astronomy Reports, 47, 59

\bibitem[{{Pavlenko}(2008)}]{pavl08co}
{Pavlenko}, Y.~V. 2008, Astronomy Reports, 52, 749

\bibitem[{{Pavlenko}(2014)}]{pavl14}
{Pavlenko}, Y.~V. 2014, Astronomy Reports, 58, 825

\bibitem[{{Pavlenko}(2019)}]{pavl19}
{Pavlenko}, Y.~V. 2019, Kinematics and Physics of Celestial Bodies, 35, 164

\bibitem[{{Peterson} {et~al.}(1993){Peterson}, {Dalle Ore}, \&
  {Kurucz}}]{pete93}
{Peterson}, R.~C., {Dalle Ore}, C.~M., \& {Kurucz}, R.~L. 1993, \apj, 404, 333

\bibitem[{{Rayner} {et~al.}(2009){Rayner}, {Cushing}, \& {Vacca}}]{rayn09}
{Rayner}, J.~T., {Cushing}, M.~C., \& {Vacca}, W.~D. 2009, \apjs, 185, 289

\bibitem[{{Rinsland} \& {Wing}(1982)}]{rins82}
{Rinsland}, C.~P. \& {Wing}, R.~F. 1982, \apj, 262, 201

\bibitem[{{Romano} {et~al.}(2017){Romano}, {Matteucci}, {Zhang},
  {Papadopoulos}, \& {Ivison}}]{roma17}
{Romano}, D., {Matteucci}, F., {Zhang}, Z.~Y., {Papadopoulos}, P.~P., \&
  {Ivison}, R.~J. 2017, \mnras, 470, 401

\bibitem[{Rothman {et~al.}(2010)Rothman, Gordon, Barber, Dothe, Gamache,
  Goldman, Perevalov, Tashkun, \& Tennyson}]{roth10}
Rothman, L., Gordon, I., Barber, R., {et~al.} 2010, Journal of Quantitative
  Spectroscopy and Radiative Transfer, 111, 2139 , xVIth Symposium on High
  Resolution Molecular Spectroscopy (HighRus-2009)

\bibitem[{Ryabchikova {et~al.}(2015)Ryabchikova, Piskunov, Kurucz, Stempels,
  Heiter, Pakhomov, \& Barklem}]{ryab15}
Ryabchikova, T., Piskunov, N., Kurucz, R.~L., {et~al.} 2015, Physica Scripta,
  90, 054005

\bibitem[{{Ryde} {et~al.}(2009){Ryde}, {Edvardsson}, {Gustafsson}, {Eriksson},
  {K{\"a}ufl}, {Siebenmorgen}, \& {Smette}}]{ryde09}
{Ryde}, N., {Edvardsson}, B., {Gustafsson}, B., {et~al.} 2009, \aap, 496, 701

\bibitem[{Sauval \& Tatum(1984)}]{84SaTaxx.partfunc}
Sauval, A.~J. \& Tatum, J.~B. 1984, ApJS, 56, 193

\bibitem[{{Schaefer} {et~al.}(2012){Schaefer}, {Lodders}, \& {Fegley}}]{scha12}
{Schaefer}, L., {Lodders}, K., \& {Fegley}, B. 2012, \apj, 755, 41

\bibitem[{{Sneden}(1991)}]{sned91}
{Sneden}, C. 1991, in IAU Symposium, Vol. 145, Evolution of Stars: the
  Photospheric Abundance Connection, ed. G.~{Michaud} \& A.~V. {Tutukov}, 235

\bibitem[{Tennyson \& Yurchenko(2012)}]{jt528}
Tennyson, J. \& Yurchenko, S.~N. 2012, MNRAS, 425, 21

\bibitem[{Tennyson {et~al.}(2016)Tennyson, Yurchenko, Al-Refaie, Barton, Chubb,
  Coles, Diamantopoulou, Gorman, Hill, Lam, Lodi, McKemmish, Na, Owens,
  Polyansky, Rivlin, Sousa-Silva, Underwood, Yachmenev, \& Zak}]{jt631}
Tennyson, J., Yurchenko, S.~N., Al-Refaie, A.~F., {et~al.} 2016, J. Mol.
  Spectrosc., 327, 73

\bibitem[{{Tinetti} {et~al.}(2018){Tinetti}, {Drossart}, {Eccleston},
  {Hartogh}, {Heske}, {Leconte}, {Micela}, {Ollivier}, {Pilbratt}, {Puig}, \&
  et~al.}]{tine18}
{Tinetti}, G., {Drossart}, P., {Eccleston}, P., {et~al.} 2018, Experimental
  Astronomy, 46, 135

\bibitem[{{Tobin} {et~al.}(2019){Tobin}, {Kemball}, \& {Gray}}]{tobi19}
{Tobin}, T.~L., {Kemball}, A.~J., \& {Gray}, M.~D. 2019, \apj, 871, 189

\bibitem[{{Vlemmings} {et~al.}(2011){Vlemmings}, {Humphreys}, \&
  {Franco-Hern{\'a}ndez}}]{vlem11}
{Vlemmings}, W.~H.~T., {Humphreys}, E.~M.~L., \& {Franco-Hern{\'a}ndez}, R.
  2011, \apj, 728, 149

\bibitem[{{Wollman} {et~al.}(1973){Wollman}, {Geballe}, {Greenberg}, {Holtz},
  \& {Rank}}]{woll73}
{Wollman}, E.~R., {Geballe}, T.~R., {Greenberg}, L.~T., {Holtz}, J.~Z., \&
  {Rank}, D.~M. 1973, \apjl, 184, L85

\bibitem[{{Yoon} {et~al.}(2014){Yoon}, {Cho}, {Kim}, {Yun}, \& {Park}}]{yoon14}
{Yoon}, D.-H., {Cho}, S.-H., {Kim}, J., {Yun}, Y.~j., \& {Park}, Y.-S. 2014,
  \apjs, 211, 15

\end{thebibliography}
\end{document}